\documentclass[
    ,final            
  ]
  {aipproc}

\layoutstyle{6x9}

\begin{document}

\title{Stars and brown dwarfs, spatial distribution, multiplicity, X-rays,
discs, and the complete mass function of the $\sigma$~Orionis cluster}


\author{Jos\'e A. Caballero}{address={Departamento de Astrof\'{\i}sica y
Ciencias de la Atm\'osfera, Facultad de F\'{\i}sica, Universidad Complutense de
Madrid, E-28040 Madrid, Spain},email={caballero@astrax.fis.ucm.es}} 

\begin{abstract}
The young $\sigma$~Orionis cluster in the Orion Belt is an incomparable site for
studying the formation and evolution of high-mass, solar-like, and low-mass
stars, brown dwarfs, and substellar objects below the deuterium burning mass
limit. 
The first version of the Mayrit catalogue was a thorough data compilation of
cluster members and candidates, which is regularly used by many authors of
different disciplines. 
I show two new applications of the catalogue and advance preliminar results on
very wide binarity and the initial mass function from 18 to 0.035\,$M_\odot$ in
$\sigma$~Orionis. 
The making-up of a new version of the Mayrit catalogue with additional useful
data is in progress.
\end{abstract}

\maketitle

\section{The {\em Mayrit} catalogue}

The constellation of Orion has pulled in the attention of astronomers and
sky-watchers since the ancient times: 
it was mentioned in the Bible, Homer's Odyssey and Virgil's Aeneid, has been
widely used in celestial navigation, and represented important characters,
deities, or tools in mythologies all over the world (it was the Babylonian
Shepherd of Anu, the Finnish V\"ain\"am\"oinen, and the Julpan, a canoe of the
Australian aborigines).
Within the constellation, the prominent Orion Belt, formed by {Alnitak}
($\zeta$~Ori), {Alnilam} ($\epsilon$~Ori), and {Mintaka} ($\delta$~Ori), is
perhaps the most famous asterism in the sky.  
Thought to be mapped in the Giza pyramid complex by the ancient Egyptians, the
Orion Belt has also been of great significance in other cultures: it was the
Scandinavian Freyja's distaff, the Iberian Tres Mar\'{\i}as, the
Northern-European Three Kings, and the Chinese lunar mansion Shen (literally
``three'' in Chinese). 

The fourth brightest star in the Orion Belt, about 2\,mag fainter than the three
main stars, is {$\sigma$~Ori}. 
The star, which is actually the hierarchical multiple Trapezium-like stellar
system that illuminates the famous {Horsehead Nebula}, has taken a great
importance in the last decade. 
Its significance lies in the very early spectral type of the hottest component
($\sigma$~Ori~A, O9.5V) and in the homonymous star cluster that surrounds the
system (Garrison 1967).  
The $\sigma$~Orionis star cluster, re-discovered due to its large number of
X-ray emitters (Wolk 1996), contains one of the best known
brown dwarf and planetary-mass object populations (B\'ejar et~al. 1999; Zapatero
Osorio et~al. 2000; Caballero et~al. 2007), and is an excellent laboratory
to study the evolution of discs and angular momenta (Reipurth et~al. 1998;
Caballero et~al. 2004; Scholz \& Eisl\"offel 2004; Oliveira et~al. 2006;
Hern\'andez et~al. 2007).
Approximate canonical ages, heliocentric distances and visual extinctions are
$\tau \sim$ 3\,Ma, $d \sim$ 385\,pc and $A_V \sim$ 0.3\,mag.

The knowledge of the whole stellar and substellar populations in a cluster in
general, and in $\sigma$~Orionis in particular, has serious implications on the
accuracy of the determination of some important parameters, such asthe slope of
the initial mass function, frequency of discs, and degree of radial
concentration.
These observational parameters are, in their turn, fundamental for the
theoretical scenarios that predict the formation of stars, brown dwarfs, and
planets.
The Mayrit catalogue of stars and brown dwarfs in the $\sigma$~Orionis cluster,
built by Caballero (2008b), was an effort to compile key data on confirmed and
candidate cluster members. 
The basis of that work was an optical-near infrared correlation between the
2MASS and DENIS catalogues in a circular area of radius 30\,arcmin contred on
the $\sigma$~Ori stellar system.
The analysis was supported by an exhaustive bibliographic search of confirmed
cluster members with signposts of youth and by additional X-ray, mid-infrared,
and astrometric data.

\section{\em Mayrit reloaded}

\begin{figure}
\includegraphics[width=0.55\textwidth]{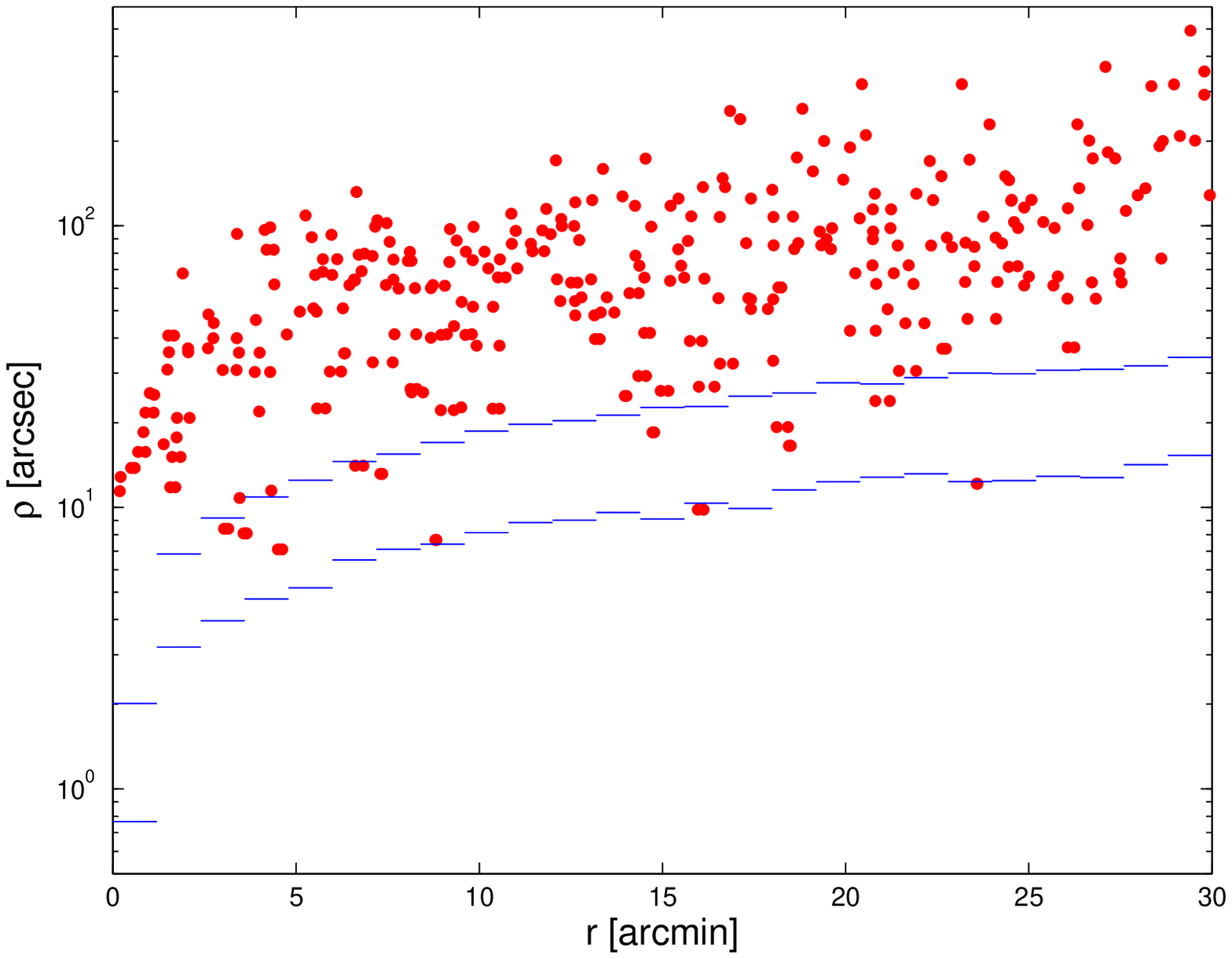}
\includegraphics[width=0.55\textwidth]{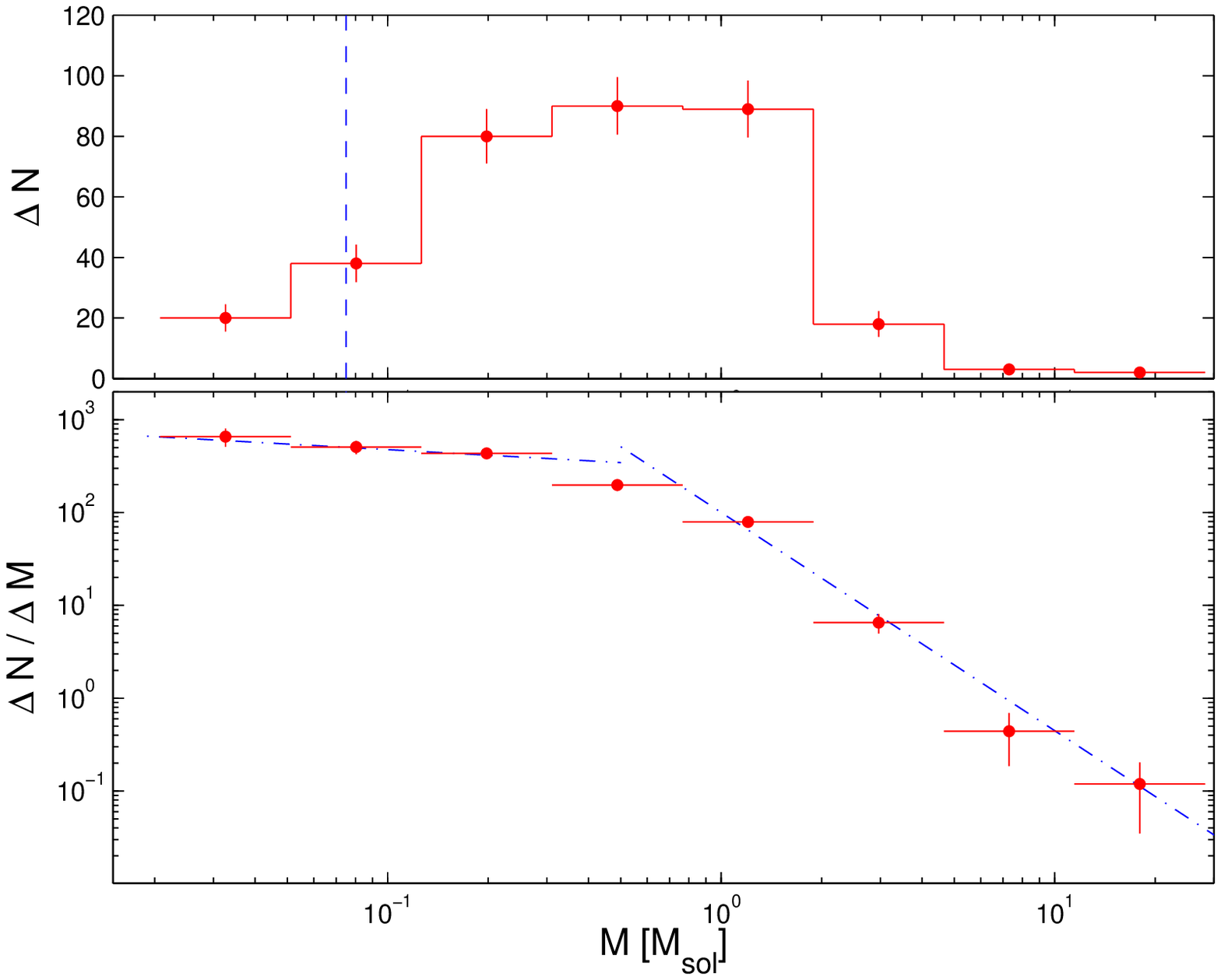}
\caption{{\em Left:} 
separation to the nearest Mayrit neighbour as a function of the angular
separation to the $\sigma$~Orionis centre.
The segmented curves denote probabilities of alignment by chance of 5 (top)
and 1\,\% (bottom). 
{\em Right:} 
mass function in the $\sigma$~Orionis cluster from 18 to 0.035\,$M_\odot$ using
the Mayrit catalogue.
{\em Top window:} cumulative number of stars and brown dwarfs per mass
interval.
The vertical dashed line line denotes the stellar/substellar boundary.
{\em Bottom window:} mass spectrum.
The diagonal dash-dotted lines indicate the slopes corresponding to $\alpha$ =
+2.35 (at high masses) and $\alpha$ = +0.3 (at low~masses).
}
\label{thefigure}
\end{figure}

The Mayrit catalogue has turned out to be a very useful tool for studying the
spatial and radial velocity distributions (Caballero 2008a, 2008c; Bouy et~al.
2008), disc frequency as a function of stellar mass (Luhman et~al. 2008), and
X-ray emission in the $\sigma$~Orionis cluster (L\'opez-Santiago \& Caballero
2008; E.~Franciosoni et~al., in~prep.).
In this proceeding, I advance some preliminar results on two new specific
topics: very wide binarity and initial mass function. 

\paragraph{Very wide binaries} 
I~have searched for very wide ($\Delta >$ 1000\,AU) systems among the Mayrit
cluster members and candidates following a simple statistical approach.
The method is based on the study of the variation of the separation to the
nearest neighbour, $\rho_{\rm min}$, with the separation to the cluster centre,
$r$ (left window of Fig.~\ref{thefigure}).  
The shorter the $\rho_{\rm min}$ between two objects is, the larger the
probability to be bound is.
However, a pair of cluster members can be separated by a short angular
separation but be located at different heliocentric distances due to a
projection effect.
I~have estimated the values of $\rho_{\rm min}$ that correspond to
probabilities of random alignments of 5 and 1\,\% for a configuration with the
same (symmetrical) radial profile as $\sigma$~Orionis. 
For that, I~have used 1000 Monte Carlo simulated distributions following the
power-law radial distribution quantified in Caballero (2008a), that corresponds
to a volume density proportional to $r^{-2}$.
Three $\sigma$~Orionis pairs of objects, listed in Table~\ref{thetable}, have
probabilities of random alignment of $\sim$1\,\%.
Interestingly, two of the ``pairs'' are actually hierarchical triple
system candidates.
Projected physical separations, of $\Delta \approx$ 2700--5100\,AU, are not rare
in the solar neighbourhood. 
Compare these values with $\Delta = 1700 \pm 300$\,AU, 
which is the separation between the $\sigma$~Orionis X-ray brown
dwarf {[SE2004]~70} and the L5$\pm$2-type, planetary-mass, object candidate
{S\,Ori~68} (Caballero et~al.~2006). 

\begin{table}
\begin{tabular}{lrr rrr rr}
\hline
	\tablehead{1}{r}{b}{Binary\\system}  		& 
	\tablehead{1}{r}{b}{Primary\\Mayrit}  		& 
	\tablehead{1}{r}{b}{Secondary\\Mayrit}  	&	 
	\tablehead{1}{r}{b}{$\rho$\\(arcsec)} 		& 
	\tablehead{1}{r}{b}{$\theta$\\(deg)}  		& 
	\tablehead{1}{r}{b}{$\Delta$\\(AU)}  		& 
	\tablehead{1}{r}{b}{$M_1$\\($M_\odot$)}  	& 
	\tablehead{1}{r}{b}{$M_2$\\($M_\odot$)}   	\\
\hline
1\tablenote{AB: [W96] 4771--899 AB, C: S\,Ori J053847.5--022711 (first proposed
	by Caballero 2005).
	The primary is a close binary ($\rho = 0.40 \pm 0.08$\,arcsec) resolved
	by Caballero (2005).
	}	& 
	528005~AB	& 530005	& 7.63$\pm$0.10	    	& 287.1$\pm$0.8	& 2700	& 1.6+?	& 0.61	  \\ 
2\tablenote{AB: OriNTT 429 AB, C: [SWW2004] 22 (first proposed here).
	The primary is a spectroscopic binary discovered by Lee et~al. (1994).}	& 
	1415279~AB	& 1416280 	& 12.18$\pm$0.10	& 4.4$\pm$0.5	& 4400 	& 1.3+?	& 0.24	  \\ 
3\tablenote{A: V507 Ori, B: S\,Ori 7 (first proposed by Caballero 2006).
	} 	& 
	397060		& 410059 	& 14.12$\pm$0.10	& 40.4$\pm$0.5	& 5100 	& 1.1	& 0.19	  \\ 
\hline
\end{tabular}
\caption{Visual systems in $\sigma$~Orionis with probability of alignment by
chance of $\sim$1\,\%} 
\label{thetable}
\end{table}

\paragraph{Initial mass function}
B\'ejar et~al. (2001), Tej et~al. (2002), Gonz\'alez-Garc\'{\i}a et~al. (2006), 
Caballero (2007), and Caballero et~al. (2007) have determined the slope of the
mass spectrum ($\Delta N / \Delta M \approx a M^{-\alpha}$) at different mass
intervals in $\sigma$~Orionis.
In right window of Fig.~\ref{thefigure}, I display the cluster (initial) mass
function in the interval 18--0.035\,$M_\odot$.
Individual masses were derived from the $J$-band magnitudes tabulated in the
Mayrit catalogue and using the 3\,Ma-old theoretical isochrones of Schaller
et~al. (1992), Baraffe et~al. (1998), and Chabrier et~al. (2000) depending on
the corresponding effective temperature.
The majority of the investigated stars have masses between 0.12 and
1.9\,$M_\odot$.
While the slope $\alpha$ of the mass spectrum in the range 18--0.7\,$M_\odot$
follows the Salpeter value ($\alpha \approx$ +2.35), it smoothly decreases down
to $\alpha \approx$ +0.3 in the mass range 0.3--0.035\,$M_\odot$. 
The transition between both slope domains is not abrupt at all, and occurs at
$M \sim$ 0.4--0.6\,$M_\odot$.
A careful derivation of the $\sigma$~Orionis initial mass function in the
interval $18 \le M/M_{\odot} \le 0.025$, acounting for a precise heliocentric
distance, incompleteness of the Mayrit catalogue, and decontamination by
late-type field dwarfs, neighbouring young stellar populations, and red quasars,
is ongoing.

\section{\em Mayrit revolutions}

I~have used the first version of the Mayrit catalogue in Caballero (2008b) as a
basis to create an updated version, Mayrit~2.0, which will be published in a
forthcoming paper. 
I have incorporated the latest spectroscopic results from Gatti et~al. (2008),
Sacco et~al. (2008), Maxted et~al. (2008), Gonz\'alez-Hern\'andez et~al. (2008),
and Caballero et~al.~(2008).  
The new catalogue contains UKIDSS $YZJHK$ magnitudes, radial velocities, and
lithium equivalent widths for about 350 $\sigma$~Orionis members and candidates.

\begin{theacknowledgments}
Partial financial support was provided by the Spanish Ministerio Educaci\'on y
Ciencia, the Comunidad Aut\'onoma de Madrid, the Universidad Complutense de
Madrid, the Spanish Virtual Observatory, and the European Social Fund.  
JAC is an {\em Investigador Juan de la Cierva} at the~UCM.
\end{theacknowledgments}

\end{document}